\documentclass[12pt,letterpaper]{article}
\usepackage[T1]{fontenc}
\usepackage[utf8]{inputenc}
\usepackage{lmodern}
\usepackage[top=1.1in,bottom=1.1in,left=1.1in,right=1.1in]{geometry}
\usepackage{amsmath,amssymb}
\usepackage{booktabs,array}
\usepackage{xcolor}
\usepackage{graphicx}
\usepackage{float}
\usepackage{caption}
\usepackage[numbers]{natbib}

\usepackage[colorlinks=true,linkcolor=blue!60!black,citecolor=green!50!black,%
            urlcolor=blue!60!black]{hyperref}
            
\usepackage{comment}
\usepackage{parskip}
\usepackage{enumitem}
\setlist{noitemsep,topsep=4pt}
\usepackage{fancyhdr}
\usepackage{lineno}

\newcommand{\pkg}[1]{\texttt{#1}}

\begin{document}

\begin{titlepage}
\centering
\vspace*{2.5cm}
{\LARGE\bfseries r2py: AI-Assisted Conversion of R Statistical Packages to Python\par}
\vspace{1.2cm}
{\large Yufei Cai\par}
{\large Jun Li$^{*}$\par}
\vspace{0.5cm}
{\normalsize Department of Applied and Computational Mathematics and Statistics\par}
{\normalsize University of Notre Dame\par}
{\normalsize $^{*}$Corresponding author: \texttt{jun.li@nd.edu}\par}
\vfill
{\normalsize\textit{\today}\par}
\end{titlepage}


\section*{Abstract}

Thousands of R packages hold statistical methods with no native Python equivalent. Runtime bridges require an R installation; hand-written ports do not scale. Translation fails silently where the languages diverge, as in transform normalization, integer width, and argument evaluation. We present \pkg{r2py}, a framework that converts an R package into a native Python library using orchestrated language-model agents under human supervision, with correctness established by numerical comparison against the original at declared tolerances. The compiled code is retained unmodified, so any divergence lies in the translation. Seven phases decompose the work for independent invocations: structural analysis fixes conversion order, every base-R construct's rendering is settled in reviewable guides before code generation, and four verification methods each expose defects their predecessors miss. Packages reaching compiled code through \texttt{.Call()} add a five-phase prologue reconstructing the R C API they use. Conversions of \pkg{KernSmooth} and \pkg{rpart} reproduce R across 518 and 846 tests.

\section{Introduction}

R~\cite{rcoreteam} is where a large share of statistical methodology is first implemented. A new estimator, test or model is typically released as an R package, and that package becomes the reference implementation the statistical and biomedical communities use, cite and build upon. \pkg{rpart}~\cite{therneau2026rpart}, the canonical implementation of classification and regression trees (CART)~\cite{breiman1984cart}, and \pkg{KernSmooth}~\cite{kernsmoothR2025}, implementing the kernel smoothing methods of Wand and Jones~\cite{wandjones1995}, are two of the fifteen packages that carry R's \emph{recommended} priority and ship with every binary distribution of R---fifteen out of roughly twenty-five thousand packages on the Comprehensive R Archive Network (CRAN)~\cite{cran}, with two thousand more on Bioconductor~\cite{gentleman2004bioconductor}. Both represent years of accumulated scrutiny. Within CRAN, 152 packages depend on or import \pkg{rpart} and 107 \pkg{KernSmooth}.

Over the same period Python has become the default environment for scientific computing, machine learning and production deployment~\cite{oliphant2007,harris2020numpy,virtanen2020scipy}. Neither \pkg{rpart} nor \pkg{KernSmooth} has a Python equivalent: \pkg{scikit-learn}'s trees~\cite{pedregosa2011sklearn} do not accept categorical predictors, and provide no surrogate splits, no cross-validated complexity table and no trees for censored or count responses; Python's kernel density estimators lack the plug-in bandwidth selectors of Wand and Jones. An analyst in Python who needs a method available only in R has one option: a runtime bridge. \pkg{rpy2}~\cite{gautier2008rpy2} and \pkg{reticulate}~\cite{ushey2023reticulate} are mature and widely used, but a bridge does not produce Python software---it requires a full R installation alongside the Python one and ties the pipeline to an environment where both runtimes are present and compatible. What is wanted is a native package, installed with \texttt{pip} and imported like any other.

Native implementations exist, and they are written by hand. \pkg{PyDESeq2}~\cite{muzellec2023pydeseq2} is a careful example: a Python reimplementation of the DESeq2 pipeline, tested against reference outputs from the R package and reported by its authors as producing results similar, though not numerically identical, to the original. Such projects show the goal is achievable. They also show what it costs: each is a bespoke effort directed at one package, and none produces a method the next project can reuse.

Large language models change the arithmetic. Models trained on code translate between languages with substantial fluency~\cite{chen2021codex,roziere2020transcoder,wang2021codet5,zhang2023survey}, and agentic decomposition extends what they can be applied to~\cite{hong2023metagpt,wu2023autogen}. Applied to code translation this has produced systems operating on whole projects: Go to Rust, validating input--output equivalence for an average of 73\% of functions in projects of up to 6,600 lines~\cite{zhang2024scalable}, and Java to Python, decomposing a repository by program analysis and translating fragments in reverse call order---an ordering we also adopt---with runtime behavior confirmed for about a quarter of them~\cite{ibrahimzada2025alphatrans}. Formal verification has been brought to bear where the target is a domain-specific language~\cite{bhatia2024}. Multilingual benchmarks now include R, but as Rosetta Code tasks of a few hundred tokens rather than packages~\cite{yan2023codetransocean}. A statistical package needs more on two axes: its whole public surface, not a validated fraction of it; and numerical agreement to a declared tolerance, which is stronger than the input--output or functional equivalence these systems establish.

This is not a matter of passing source files to a model. A package is a dependency graph dictating what must be translated before what; its numerically intensive work sits in C or Fortran written against R's own runtime; a package of any size exceeds a single model context, so the work must be divided among independent, stochastic invocations sharing no memory of decisions taken elsewhere; and establishing correctness means comparing numerical output across the whole public surface, including the error paths ordinary use does not reach. Hardest of all, R and Python diverge in ways inspection of correct-looking code does not reveal. R's \texttt{fft(inverse=TRUE)} returns an unnormalized inverse transform while NumPy's \texttt{ifft} divides by $N$; R's \texttt{var()} applies Bessel's correction while \texttt{np.var} does not; Fortran \texttt{INTEGER} is 32 bits where NumPy defaults to 64. Each produces plausible numbers that are wrong, with no exception raised. Beyond these, R has constructs with no Python counterpart at all---S3 dispatch, model formulas, environments---whose translation is a design decision rather than a lookup.

We present \pkg{r2py}, a framework that converts an R package into a native Python package using large language models under human supervision, with correctness established by numerical comparison against the original. It proceeds as an ordered sequence of phases, each executed by orchestrating \emph{skills} that dispatch specialized \emph{sub-agents} with independent model contexts: analyzing structure before translating anything, fixing the Python rendering of every base-R construct in a written guide before any code is generated, converting in dependency order, reusing the original compiled code, and validating against the original both function by function and end to end.

We applied it to \pkg{KernSmooth} and \pkg{rpart}, chosen to differ in the one respect that governs the work: the first reaches its compiled core through R's array-based \texttt{.Fortran()} interface, the second through \texttt{.Call()}, which exchanges R's own internal objects. A seven-phase core carried both; \pkg{rpart} additionally required a five-phase prologue rebuilding the portion of R's C API it uses. The resulting packages reproduce their originals across 518 and 846 tests benchmarked against R, and are distributed on PyPI.

\section{Results}

\subsection{The \pkg{r2py} framework}
\pkg{r2py} organizes the conversion of an R package into seven phases (Fig.~\ref{fig:pipeline}), each executed by large language models under human supervision and each addressing a specific way in which a translation can go wrong.

\begin{figure}[H]
\centering
\includegraphics[width=\textwidth]{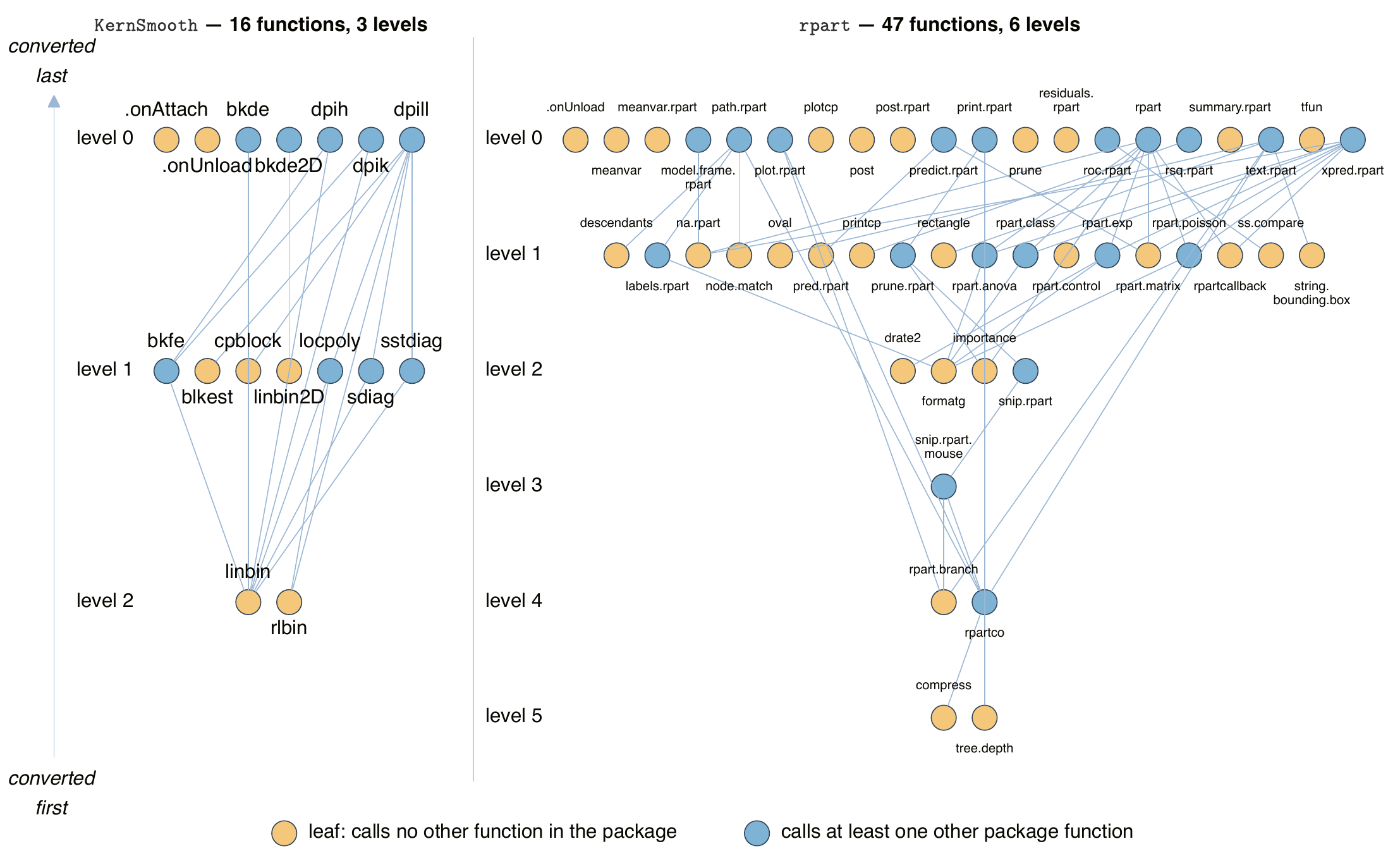}
\caption{\textbf{The dependency graph that fixes conversion order.} Each node is a function defined by the package; an edge joins a function to one it calls. Functions are assigned a level by longest path from any root, and conversion proceeds from the deepest level upward, so that when an agent translates a caller the Python interfaces of everything it calls already exist. Leaves---functions calling no other function in the package---carry no such prerequisite, and can be translated in any order. \pkg{KernSmooth}'s 16 functions resolve into three levels and \pkg{rpart}'s 47 into six. Both graphs are wide and shallow: 7 of \pkg{KernSmooth}'s functions and 28 of \pkg{rpart}'s are leaves, so much of each conversion carries no ordering constraint at all, and only five of \pkg{rpart}'s 47 functions lie deeper than level two. The graph is the output of phase~1 and is deposited with the software.}
\label{fig:dag}
\end{figure}

The order is not editorial: each phase depends on what the previous one established. Structural analysis must come first, because the dependency graph it recovers (Fig.~\ref{fig:dag}) determines the order in which functions can be translated at all. Build infrastructure precedes translation so that each converted function is executable when written. Cataloging every base-R construct and fixing its Python rendering precedes generation, so that a construct is rendered identically wherever it appears rather than decided afresh by each independent invocation. Only then is the R layer converted, function by function in dependency order, and assembled into modules. The three verification phases run last and in sequence, because reading, executing and testing expose different classes of defect.

A design decision shapes the rest of the framework. Statistical R packages seldom perform their numerical work in R, and reimplementing a compiled routine in interpreted Python would forfeit the performance that motivated compiling it. \pkg{r2py} therefore retains the original compiled source unmodified and translates only the R layer that calls it. This also has a consequence the framework depends on: because the computational core executing under Python is byte-identical to the one executing under R, any numerical divergence must originate in the translated layer---small enough to audit exhaustively, whereas a fault inside reimplemented compiled code would surface as a plausible wrong number.

What that decision costs differs between packages---not by size or complexity, but by the interface through which R reaches its compiled code, and we encountered both cases.

\pkg{KernSmooth} delegates its computation to Fortran~77 reached through R's \texttt{.Fortran()} interface, which passes plain, C-compatible arrays by reference. Both languages agree on the binary representation of such an array, so the retained Fortran compiles into the Python package and is called with NumPy buffers directly (Methods). The compiled-code decision costs nothing beyond the build configuration of phase~2, and \pkg{KernSmooth} passes through the seven phases unchanged.

\pkg{rpart} reaches its C code through \texttt{.Call()}, and this changes the problem entirely. \texttt{.Call()} does not pass arrays; it passes \texttt{SEXP} values, pointers to R's own internal objects. C code written against this interface allocates R objects, registers them with R's collector, reads them through R's accessor macros and signals errors through R's own mechanism---all provided by \texttt{libR.so} at run time. Python cannot construct such objects, and shipping an R installation with every Python wheel would reinstate exactly the dependency the conversion exists to remove.

This is not a peculiarity of \pkg{rpart}. R's own extension documentation directs authors toward \texttt{.Call()} wherever compiled code exchanges anything beyond plain numeric arrays~\cite{rexts}, and the obstacle extends well past the packages that call it directly: \pkg{Rcpp}~\cite{eddelbuettel2011rcpp}, on which more than a thousand CRAN packages are built, generates \texttt{.Call()} entry points over the same \texttt{SEXP} representation, so its dependants inherit the same barrier. Any framework intending to convert more than the easiest packages has to solve it.

We solve it by reconstructing, as standalone code, exactly the portion of R's C API the target package uses, so that its original C sources compile and run with no R present. This adds a \emph{prologue} of five phases ahead of the seven-phase core: analyzing the compiled source's dependencies, inventorying the R C API symbols it references, writing a blueprint for each, generating the replacement headers, and producing entry-point wrappers callable from Python. The core itself is unchanged: \pkg{rpart} then passes through the same seven phases \pkg{KernSmooth} did. Section~\ref{sec:sexp} describes the reconstruction.

Between them the two cases span what a further package is likely to present. Reached through \texttt{.Fortran()} or \texttt{.C()}, its compiled code passes arrays and needs no prologue; reached through \texttt{.Call()}, or through \pkg{Rcpp} over the same representation, it needs the reconstruction---and which case it falls into is readable from its source beforehand.

\subsection{How the framework is executed}
\begin{figure}[H]
\centering
\includegraphics[width=\textwidth]{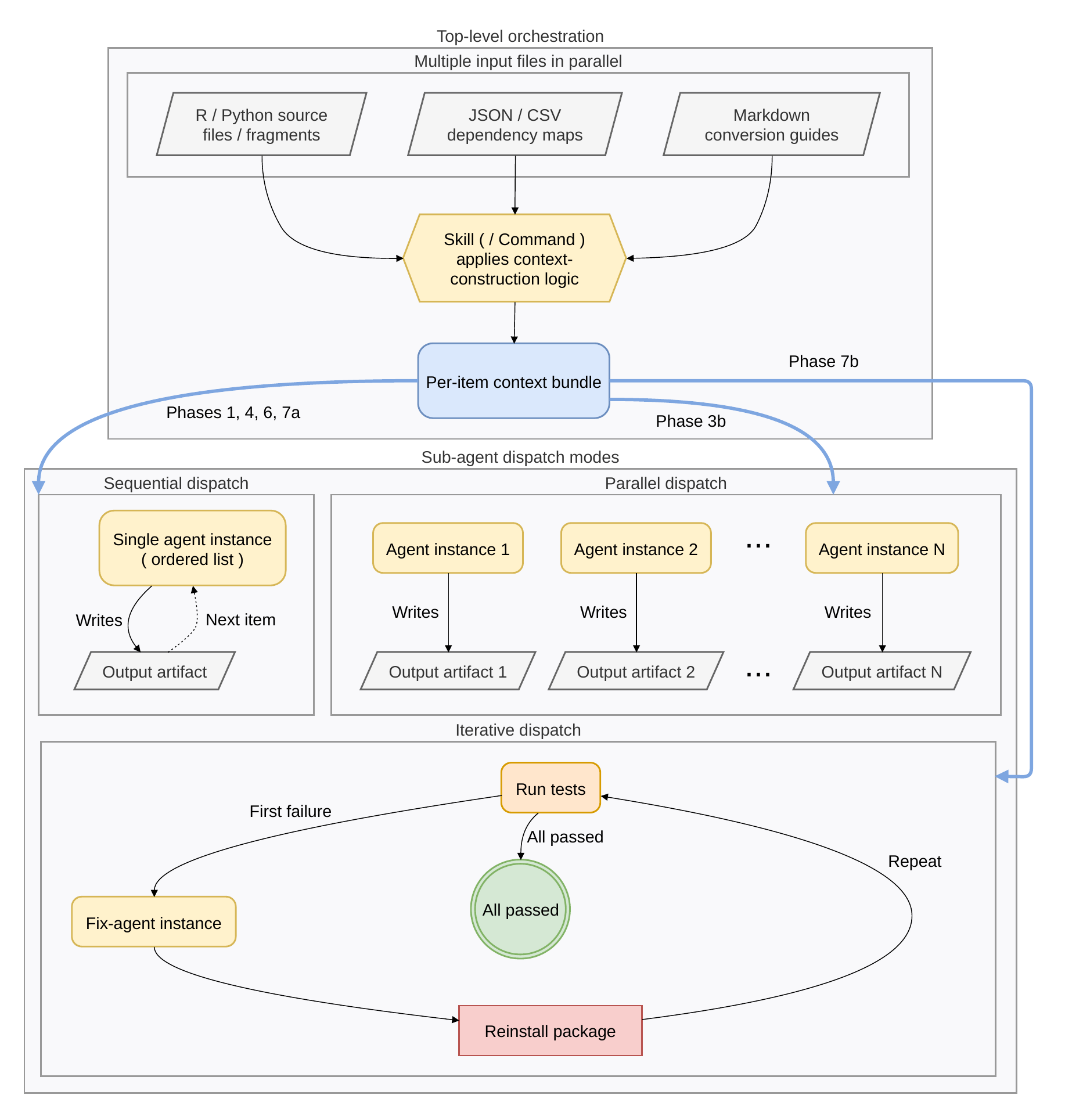}
\caption{\textbf{Skill--sub-agent orchestration.} A skill assembles per-item context bundles from shared inputs and dispatches sub-agents in one of three modes. The mode is determined by the dependency structure of the work rather than chosen for convenience. Sequential dispatch is used where an invocation must read the output of an earlier one; parallel dispatch where invocations are mutually independent; iterative dispatch where a fix-and-recheck cycle runs until a stopping criterion is met.}
\label{fig:orchestration}
\end{figure}

Each phase except two---build infrastructure and the equivalence audit---is carried out by a \emph{skill}---an orchestrator that assembles the inputs a unit of work requires and dispatches it---and by the \emph{sub-agents} it invokes, each a separate language-model invocation with its own context (Fig.~\ref{fig:orchestration}). We instantiated the framework in Claude Code (Anthropic)~\cite{claudecode}, but the decomposition does not depend on that choice and transfers to any environment supporting multi-agent orchestration with file-system access.

The work is divided into units small enough to be carried out reliably---one source file, one construct, one function, one failing test---and almost every artifact is model-generated: the structural analysis, the construct catalog and its guides, the Python functions, and for \pkg{rpart} the symbol inventory, the 51 replacement headers and the entry-point wrappers, together with the test suites and the repairs applied in the bug-resolution loop. What the framework contributes is not the generation but the decomposition: deciding what each invocation should see, in what order they must occur, and what has to be settled before any begins.

The decomposition rests on a principle: every cross-cutting translation decision is settled once, in a machine-readable artifact, before any code is generated. R's base library does not map one-to-one onto Python, and the correct translation is often context-dependent---\texttt{as.integer} is called at 33 sites in \pkg{rpart} across at least half a dozen structurally distinct contexts, and a strategy correct in one is not automatically correct in another. Because each agent runs as an independent, stochastic invocation with no memory of decisions taken elsewhere, resolving such questions case by case offers no way of detecting inconsistency short of exhaustive pairwise comparison. The framework therefore records every call site of every construct---509 across 46 constructs in \pkg{KernSmooth}, 1,298 across 172 in \pkg{rpart}---and generates one authoritative guide per construct. An implicit, per-invocation judgment becomes an explicit artifact a reviewer can correct in one place before the error propagates into many functions.

Human effort is not eliminated but redirected. It goes into specifying each phase, reviewing its output before the next begins, and making the decisions that are not mechanical (Methods). A manual cross-check of \pkg{rpart}'s 35 structural-analysis files caught three misclassified call sites that would otherwise have propagated into the conversion order. In each case the model performed the work and a human decided whether it was right.

\subsection{Validating the conversion}
\label{sec:validation}
The conversion targets faithfulness in a strict sense: we do not reorganize the package, redesign its public interface, or improve its algorithms, and the Python package defines the same functions with the same arguments and returned structures. This is what makes function-level comparison possible at all---each converted function has a counterpart in R taking the same arguments, and can be checked against its original in isolation rather than only through end-to-end output.

The correctness criterion is numerical equivalence to the original implementation. This is a stronger and narrower requirement than the functional or structural equivalence used in general-purpose code translation, where a translation succeeds if it computes the same thing in some recognizable sense. Here a translation succeeds only if it returns the same numbers, to a tolerance declared at the point of comparison and checked against R actually running (Methods).

A suite only finds defects on paths it exercises. The generated tests are written with that in mind, in three categories---positive, negative and boundary (Methods)---of which the boundary cases matter most, since the edges of the input domain are where translation defects concentrate. But no suite closes the gap by itself. Four verification methods were applied in sequence---the two halves of the phase-5 audit, the package's own tests in phase~6, then the generated suite in phase~7---and their differential yield is itself a result.

A systematic static audit, in which agents compared each translated module against its R source without executing either, found real defects across every subsystem. Running the assembled package on realistic workflows then exposed roughly sixteen further issues the static reading had been structurally unable to see---violations of runtime semantics rather than of translated logic. Translating the package's own historical regression tests surfaced further defects in six of the fourteen scripts, none of which any earlier method had found; the generated suite, run last, added more still. Each method ran against a package already corrected by its predecessors, so the sequence measures what each reaches beyond the last (Table~\ref{tab:defects}).

The clearest instance is a defect that only the third method reached. A control-parameter default was computed as 21 where R computes 20, so the implementations diverge only in trees deep enough to reach a node of exactly twenty observations; downstream, the effect is a discrepancy in cross-validated error indistinguishable from floating-point noise. Neither the static audit nor the workflow execution that preceded it exercised a tree that deep (Methods).

What the defects in Table~\ref{tab:defects} have in common is that they do not announce themselves. Most returned plausible numbers; one corrupted the heap several calls after the offending assignment. None would have been caught by a suite checking only that the code runs, and several would have survived any review that did not execute it. This is the concrete form of the risk that motivates the framework.

Both conversions reproduce their R originals at every asserted tolerance (Table~\ref{tab:twopkg}); the median assertion bounds agreement at $10^{-6}$ in \pkg{r2py\_kernsmooth} and $10^{-7}$ in \pkg{r2py\_rpart}. Two exceptions among \pkg{rpart}'s 846 tests record a divergence that cannot be closed rather than one that has not been: R's \texttt{print} and \texttt{summary} echo the call as the user typed it, reconstructed from the unevaluated expression R captures at the call site, which a Python object retaining only evaluated arguments cannot reproduce. A further 94 differences were cataloged in which both languages reject the same input but describe the rejection differently; these are recorded, not failed.

These results bound what has been established as well as what has been achieved: the measured agreement is a lower bound on true correctness, holding at the tolerance asserted and only along the paths the suites happen to exercise. The layered sequence provides not a guarantee but a succession of independent opportunities for a defect to be caught.

\subsection{Reconstructing R's C API}
\label{sec:sexp}

Reconstructing R's C API means supplying, in standalone code, the runtime services the package's C source expects: the object representation, the allocators, the accessors, and the error and output routines. The work is bounded by what the package actually uses---for \pkg{rpart}, 235 reference sites across 51 distinct identifiers, established by static analysis before any of it was written (Methods).

\begin{figure}[H]
\centering
\includegraphics[width=\textwidth,keepaspectratio]{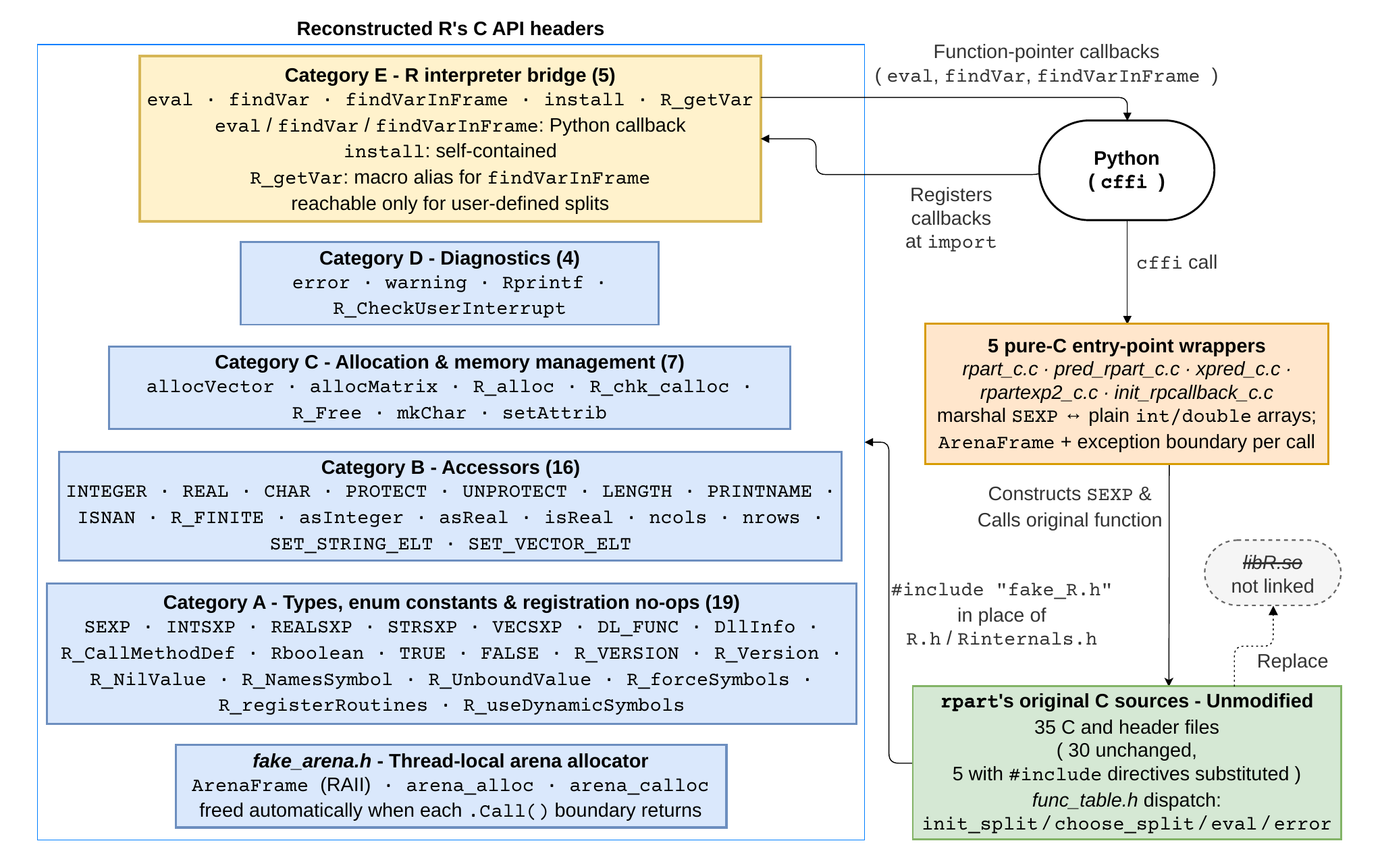}
\caption{\textbf{Reconstructing R's C API as standalone headers.} The prologue replaces R's C API with self-contained implementations of exactly those symbols the target package references, so that the package's original compiled sources build and execute without an R runtime. For \pkg{rpart} the 51 symbols divide into five categories of sharply differing difficulty: types and registration no-ops (19), accessors (16), allocation and memory management (7), diagnostics (4), and operations on R's interpreter (5). Only the last category reaches beyond what standalone C++ can provide. Four of the five are resolved by callbacks into Python and the fifth by a compatibility macro that expands into two of those four; of the four, only expression evaluation genuinely requires an interpreter. In \pkg{rpart} the category as a whole is reachable only when the user supplies a custom splitting rule.}
\label{fig:fakeapi}
\end{figure}

The 51 items are not equally difficult, and the distribution is what makes the approach viable (Fig.~\ref{fig:fakeapi}): forty-six have direct standalone equivalents. Of the five that are operations on R's interpreter, one is never linked---a compatibility macro expands it into the two environment-lookup routines---and the other four are reached only on the user-defined-split path, where Python registers callbacks for them as that path initializes. Only expression evaluation truly requires an interpreter; the rest reduce to a pointer table, a lookup in it, and an unreachable stub. This residue is the honest limit of the approach, and its size is why the approach works at all: in \pkg{rpart}, the bridge is reachable only when a user supplies a custom splitting rule, and all four built-in methods execute without entering it.

R distinguishes transient allocations from persistent objects that outlive a call, and the reconstruction preserves the distinction (Methods). A consequence is that \texttt{PROTECT} and \texttt{UNPROTECT}, the protocol that exists to shield objects from garbage collection, have nothing to shield them from and reduce to no-ops.

The reconstruction is header-only, one self-contained header per API item, and is written in C++ because the callback pointers and the sentinels standing in for R's \texttt{NULL} must resolve to a single instance across all 37 translation units (31 package sources, five entry-point wrappers and one helper file). C++ inline variables deliver that from a header alone; C would need a coordinating source file owning all of them---precisely the ordering-sensitive shared artifact that generating one item at a time avoids.

Whether such a reconstruction is faithful is not settled by inspecting it. The first evidence is how little of the original had to change: of the 35 files, 30 required no modification whatsoever and 5 only a substitution of include directives. Beyond that a single anonymous struct declaration, legal in C but not in C++, needed a tag name; no referenced symbol had to be stubbed out. The second is the numerical result, which the validation of Section~\ref{sec:validation} settles.

The difficulties encountered were not statistical or algorithmic but properties of separate compilation, and they indicate where the risk in such a reconstruction lies. An audit of the generated headers found seven defects, most of them include-guard mismatches or linkage conflicts (Methods). The most consequential was of a different kind: the four function-pointer variables underlying the interpreter bridge had been declared \texttt{static} at namespace scope, which in C++ gives each translation unit a private copy, so a callback registered from Python would have updated one unit's pointer and left every other holding a null. The defect compiles cleanly, links cleanly, and fails only when the callback path is first exercised---in \pkg{rpart}, only when a user supplies a custom splitting rule. The prologue is bounded and enumerable before it is begun, but its risks are of this kind.

\section{Discussion}

The two conversions suggest that the barrier separating statistical software from the Python ecosystem is more tractable than its reputation, and that what makes it tractable is not the capability of the underlying model but the discipline imposed around it. Almost every artifact was model-generated; almost every decision about what each invocation should see, and in what order, was not.

It is worth stating plainly what this contribution is and is not. It is not the conversion of two packages, important though those are to the people who use them; neither is it a turnkey procedure that converts an arbitrary R package without intervention. Every phase reported here was specified, reviewed and in places corrected by us, and both target packages were chosen because they depend on nothing beyond base R. What we offer is a decomposition---an ordered set of phases, each with a defined artifact and a defined check---whose cost on a further package can be estimated before the work begins. Packages carrying contributed dependencies, or exposing a wider compiled-code boundary, will need components this framework does not yet supply. We report it as a demonstration that conversions of this kind can be carried out systematically rather than bespoke, and as a structure for the ones that follow.

Faithfulness, as we have defined it, reaches further than preserving an interface: the harder part concerns behavior the original gets wrong. During validation we identified a genuine fault in \pkg{rpart}'s own C source (Methods). The file is byte-identical in the original and in our port, so the defect is reproduced exactly, and we have deliberately not fixed it: an analysis published using R's \pkg{rpart} should reproduce under \pkg{r2py\_rpart}, and a port that silently improved on its original would produce discrepancies users would rationally attribute to the port rather than the correction. The defensible position is to document such defects explicitly rather than propagate them silently. A reimplementation may be better than its source; a port must be the same. The same applies to credit: the results a port produces are its original's, and work relying on it should cite both.

Of the four verification methods reported above, the one we would most strongly recommend is the one easiest to overlook: translating the regression suite the source package already ships. Such suites reach states that neither a reviewer nor a generated suite has reason to construct. The control-parameter defect was reachable only that way, having survived both a systematic audit and execution on realistic multi-step workflows. We expect this to hold for AI-assisted porting generally.

The framework's artifacts are reusable across conversions, but only partly. The translation guides document how base-R constructs render in Python, and such constructs recur across R packages generally---but the two catalogs overlap without nesting, and each conversion enlarged the catalog rather than drawing on it unchanged (Methods). Because each guide enumerates the usage patterns observed in \emph{its} source package, a construct already cataloged may still need extending when a new package exercises it differently. The reconstructed headers are reusable in the same qualified sense, making the prologue closer to a one-time cost than a cost recurring with every package.

Three limitations bound what we have shown. A package carrying a tree of contributed CRAN dependencies would require the framework to be applied recursively, or each dependency mapped onto a Python equivalent, and we have not characterized what that costs; much of the biostatistical software that motivates this work, including a large part of Bioconductor, is of that kind. The compiled and interpreted layers could be developed parallelly only because \pkg{rpart} reaches its C layer through five fixed-signature entry points; the boundary still grew to fifteen functions once a user-defined split calls back into Python (narrowness where R calls in does not imply narrowness where C calls back), though a prospective conversion can at least measure its own before assuming it does. Some R-specific behavior remains out of reach: the printed call echo behind our two expected failures, one interactive routine, and formula support covering the common forms rather than R's full model-frame semantics.

A different kind of limit applies to the evidence itself. Both conversions were carried out with the complete framework in place, so the marginal contribution of any individual phase is not isolated here. This matters most for the phase producing no code at all---the language-dependency catalog---whose value we argue from the consistency requirement rather than demonstrate by measurement.

The immediate extensions are to extract the reconstructed API headers as an independent library and to attempt a package with genuine CRAN dependencies, which is where the framework's limits will next be found. We expect improving models to reduce the human effort each phase requires rather than to remove the phases themselves, since the decomposition follows from the structure of the problem rather than from any current limitation of the models. The result we would emphasize is not that two packages were converted but that the cost of converting a third is largely predictable in advance from a static scan of its source.

\section{Methods}

Results report what the framework does and why; this section records how each phase was carried out, in enough detail to reproduce it. It is organized by layer rather than by phase number---the R layer first, then the compiled layer and the interface to it, then validation and distribution---because the machinery within a layer is easier to follow together than split across the order in which it ran. The skill and sub-agent specifications themselves are Markdown documents with YAML front matter declaring a name and description, followed by a structured natural-language body giving execution steps and an output schema. They reside in \texttt{.claude/commands/} and \texttt{.claude/agents/} in the project tree and are reproduced in full in the Supplementary Information. The conversions themselves are recorded in two Supplementary Notes---Note~1 for \pkg{KernSmooth}, Note~2 for \pkg{rpart}---written contemporaneously with the work rather than composed afterwards. Each documents every phase as it was carried out, including the diagnostic paths taken through defects that were subsequently fixed, so that the process reported here can be audited rather than taken on trust.

\subsection*{Target packages}

\pkg{KernSmooth}~\cite{wandjones1995} implements the kernel smoothing methods of Wand and Jones: binned kernel density estimation in one and two dimensions, kernel functional estimation, local polynomial regression, and---most distinctively---plug-in bandwidth selection~\cite{sheatherjones1991,ruppertsheatherwand1995}, computed over binned data for speed~\cite{wand1994}. It was written by M.~P.~Wand, with the R port maintained by Brian Ripley. Here too the gap is specific rather than total: Python has kernel density estimators, but the direct plug-in bandwidth selectors that make Wand and Jones's methods the reference for bandwidth choice, and local polynomial regression with binning, are absent or only partially available.

\pkg{KernSmooth} version~2.23-26 (license: Unlimited) defines 16 functions in a single R source file, 7 of which are exported. Its computational work is delegated to 11 fixed-form Fortran~77 files exposing 8 subroutines, registered through \texttt{useDynLib(KernSmooth,\ .registration = TRUE,\ .fixes = "F\_")}. Three LINPACK routines are included for LU decomposition. Two further LINPACK routines providing QR decomposition (\texttt{dqrdc}, \texttt{dqrsl}) are called by the package but satisfied under R from R's own internal LINPACK copy; because OpenBLAS does not supply them, they were obtained from Netlib and added to the Python package's sources, bringing its compiled sources to 13 files.

\pkg{rpart}~\cite{therneau2026rpart} implements the CART methodology of Breiman, Friedman, Olshen and Stone~\cite{breiman1984cart}: recursive partitioning for continuous, categorical, count, and censored responses, with surrogate splits for missing predictors, cost-complexity pruning, and cross-validated selection of tree size. It was written by Terry Therneau and Beth Atkinson, with the original R port by Brian Ripley, and it is the canonical implementation against which other decision-tree software is compared. Python has decision trees, but not these: \pkg{scikit-learn}'s implementations provide neither surrogate splits, nor the cross-validated complexity table, nor trees for censored or count responses, nor user-supplied splitting rules. A Python user who needs any of those has no native option. The difference in categorical handling is one of model class rather than convenience. \pkg{rpart} may split a factor on an arbitrary subset of its levels, recording each level's assignment in a \texttt{csplit} matrix; the package's own vignette enumerates all $2^5=32$ combinations of a six-level predictor in a worked example. \pkg{scikit-learn} describes its trees as an optimized version of the CART algorithm but states that its decision-tree implementation ``does not support categorical variables for now'', so factors must be encoded numerically before fitting---and an encoding recovers the same partition only indirectly: a chain of indicator splits can isolate a subset but spends one level of depth per level of the factor and is found, if at all, by a greedy search that evaluates the indicators one at a time, while ordering the levels by mean response---the shortcut \pkg{rpart} itself uses for two-class, anova and Poisson outcomes---recovers the optimal subset split only for those outcomes, not for the multi-class case where the exhaustive search is required.

\pkg{rpart} version~4.1.27 (GPL-2 $\vert$ GPL-3) defines 47 functions across 36 R source files, with 35 C source and header files. Its C layer is organized around a dispatch table mapping a method integer to four function-pointer slots---node initialization, split selection, evaluation, and error computation---which keeps the recursive partitioning logic method-agnostic; method integer 4 denotes a user-supplied splitting rule implemented as a callback into the R interpreter. The R layer reaches the C layer through five \texttt{.Call()} entry points: \texttt{rpart}, \texttt{pred\_rpart}, \texttt{xpred}, \texttt{rpartexp2}, and \texttt{init\_rpcallback}.

\subsection*{Structural analysis and conversion ordering}

One sub-agent was dispatched per R source file. Each enumerated every function definition, including nested closures, and classified every call within each function body into exactly one of three categories: \emph{language dependencies} (base R and its standard library), \emph{internal dependencies} (functions defined elsewhere in the same package), and \emph{external dependencies} (\texttt{.Fortran()}, \texttt{.C()}, \texttt{.Call()}, or \texttt{.External()} crossings into compiled code). Output was one JSON object per file, keyed by function name, with the three dependency lists as values.

Internal dependencies were assembled into a directed acyclic graph and each function assigned a level by longest path from any root; the resulting table records, for each function, its level, immediate callers, and immediate callees. \pkg{KernSmooth} resolved into three levels, with the one-dimensional linear binning routine called by seven distinct callers. \pkg{rpart} resolved into six levels, 28 of its 47 functions making no internal calls at all. Conversion order was read from this table, deepest level first. For \pkg{rpart} the analysis was cross-checked by hand, which corrected three misclassified call sites before they could propagate into the conversion order.

\subsection*{Language-dependency catalog and translation guides}

A second per-file pass recorded every individual call site of every language dependency identified in the structural JSON, capturing the dependency name, enclosing function, line number, and complete call expression including multi-line calls. Per-file CSVs were merged and sorted by dependency name so that all sites for a given construct are contiguous. One CSV field required correction: R string literals embedded in call expressions had been written with bare rather than doubled double-quote characters, breaking RFC~4180 parsing; the file was rewritten using a conforming CSV writer.

The catalog held 46 distinct constructs over 509 call sites for \pkg{KernSmooth} and 172 over 1,298 for \pkg{rpart}; the two sets overlap without nesting, 34 constructs being common to both, 12 peculiar to \pkg{KernSmooth} and 138 to \pkg{rpart}. One sub-agent was then dispatched per unique construct, in parallel, each receiving only the CSV rows for its assigned construct. Agents read the corresponding source lines for context and consulted R and Python documentation where behavior was non-obvious. Each produced a Markdown guide with four parts: an overview of the construct and its translation hazards; a contextual usage analysis enumerating every call site grouped by usage pattern; a general conversion strategy; and step-by-step examples pairing R code with its Python translation for each distinct pattern. Construct names containing characters not permitted in filenames were encoded; two guides are dot-prefixed and therefore invisible to shell globbing, which required directory listing rather than globbing to verify the expected count.

\subsection*{Function conversion and package assembly}

Conversion was performed one function per sub-agent invocation, sequentially in the topological order established above. Each agent received the R source for its function, the function's structural JSON entry, the folder of translation guides, and the dependency-level table. Output was a JSON artifact with three fields: the import statements required, the complete Python signature with type annotations, and the function body as a list of lines. This encoding, rather than direct emission of Python source, allows an agent converting a caller to read its callees' signatures and return types mechanically.

R constructs without direct Python equivalents were handled uniformly: named lists became dictionaries; \texttt{missing()} became a module-level sentinel object tested by identity; S3 class assignment became a reserved dictionary key; environments became dictionaries; 1-based indices were converted at every crossing; and matrices passed to compiled code were laid out in column-major order.

A separate skill assembled the artifacts into modules. Assembly corrected three classes of systematic defect: absolute import paths were rewritten as relative imports; redundant cross-imports between functions sharing a module were removed; and duplicate third-party imports were merged, with the final import block ordered standard library, third party, local. For \pkg{rpart}, 26 of 36 output filenames contained dots inherited from R naming conventions and were renamed with underscores to be importable. Syntax was confirmed by parsing each assembled module. Because \texttt{meson-python} does not auto-discover Python sources, every module must be enumerated explicitly in the build definition; this is a standing maintenance requirement whenever modules are added.

\subsection*{Equivalence audit and test porting}

The assembled package was audited in two passes. In the first, one sub-agent per module compared the translated Python against its R source without executing either, checking that control flow, error conditions and returned structures corresponded, and reporting discrepancies for review; findings were approved before any source file was changed, under a standing constraint that the retained compiled sources were never to be altered. In the second, the package was executed on realistic multi-step workflows, which exposes violations of runtime semantics that static reading cannot reach.

Where the source package ships a regression suite, each script was translated by one sub-agent and run against the Python package. \pkg{rpart} ships fourteen such scripts and \pkg{KernSmooth} two. Because these suites depend on datasets and on printed output, reference values and datasets were captured once from a live R session into checked-in files rather than obtained through \pkg{rpy2} at test time.

\subsection*{Reconstruction of R's C API}

The inventory, the category breakdown, and the defects found in the generated headers are reported in Results; this section records the representations and mechanisms those results depend on.

\texttt{SEXP} is a pointer to a structure carrying a type tag, a length, row and column counts, and a data pointer. Implementation blueprints were written for the 51 items sequentially rather than in parallel, because foundational definitions---the structure layout, the type constants, the allocator---must be fixed before dependent items can reference them; headers were then generated in the same order and assembled behind a single master include.

The arena serving transient allocations is thread-local. Each entry-point wrapper declares an arena frame as its first local variable, so that the frame's destructor releases every block allocated during the call on all exit paths, including exceptional ones. Persistent objects---\texttt{SEXP} nodes and their data buffers---are allocated on the heap and freed explicitly by the caller after the boundary returns.

Two choices could not be derived mechanically from the item inventory. The faked R version constant had to satisfy two opposed preprocessor guards simultaneously---one requiring at least 2.16.0, another requiring less than 4.5.0---and was set to a value in that window. Of the five interpreter operations, one variable-lookup entry point resolves to a compatibility macro expanding to two other items in the same category; the other four---expression evaluation, symbol interning and the two environment-lookup routines---are resolved through function pointers, which Python registers when the callback path is initialized rather than at import, so a fit that never supplies a custom splitting rule never registers them. Only expression evaluation requires an interpreter; the interning callback merely returns a stable handle for each name, for which the headers also retain a self-contained implementation, and the two lookups reduce to a table indexed by those handles, one of which \pkg{rpart} never reaches.

Five entry-point wrappers, compiled as C++ but exposing C linkage, translate between the \texttt{SEXP}-based internal interface and a plain-array interface. Each includes the master header outside its \texttt{extern "C"} block, declares an arena frame first, and implements an exception boundary catching the package's own error type, allocation failure, standard exceptions, and any remaining exception, writing a null-terminated message to a caller-supplied buffer. Environments and expressions, which have no plain-array representation, are passed as opaque handles; these are cast directly rather than re-wrapped, since the identity of the pointer is what the Python side uses as a registry key. A further helper file exports routines for constructing and releasing \texttt{SEXP} nodes that wrap NumPy buffers without copying.

Local wrapper macros defined in the package's own header were excluded from the symbol inventory even where they wrap R API calls, since those travel with the source. The sentinel objects standing in for R's \texttt{NULL} and its symbol table use a construction structurally similar to the function-pointer variables---file-local pointers initialized by inline factory functions containing function-local statics---which was examined and found safe, since every translation unit ultimately references one shared instance and pointer comparison remains valid.

Migration of the original sources to the reconstructed headers was dispatched in parallel across all 35 files, since the edits are independent.

\subsection*{Foreign-function interface semantics}

An \pkg{f2py}~\cite{peterson2009f2py}-wrapped Fortran subroutine returns \texttt{None} unconditionally; output arrays are modified in place through the pointer passed in, and the caller reads results from the pre-allocated arrays rather than from a return value. Scalar outputs require particular care, since a Python float is immutable and cannot be written back: either a length-1 array is passed, or the subroutine is given an explicit output intent. For the three scalar outputs of the block-estimation routine, \pkg{f2py} had inferred pass-by-value from the Fortran declaration, so the computed values never propagated; this was corrected by generating an explicit signature file with output intent declared and building from that file rather than from the bare sources.

Two further behaviors of NumPy~2.4.3 affected the wrappers. Dimension scalars declared in a Fortran signature are absorbed out of the required positional arguments and inferred from the shapes of the arrays passed, so wrappers retaining them at their original positions misalign every subsequent argument; removing them reduced the local-polynomial call from 19 to 16 positional arguments, with corresponding reductions elsewhere. Separately, parameters declared two-dimensional in Fortran are shape-checked strictly, so arrays must be allocated in Fortran order rather than flattened before the call.

For \pkg{rpart}, the compiled code is loaded as a shared library through \pkg{cffi}~\cite{cffi} in application-binary-interface (ABI) mode rather than built as a Python extension module. NumPy buffers are passed by taking a buffer view; callbacks are registered as \pkg{cffi} callback objects; and both the buffer objects and the views derived from them are retained in a keep-alive list for the duration of a call, since neither is owned by the C side.

\subsection*{Build system and distribution}

Both packages use \texttt{meson-python} as the Python Enhancement Proposal (PEP)~517 build backend. \pkg{r2py\_kernsmooth} builds a Python extension module from an \pkg{f2py} custom target, with NumPy's C and \pkg{f2py} include directories resolved at configure time. BLAS is located through a five-stage chain: \texttt{pkg-config}, then library searches for \texttt{openblaso}, \texttt{openblas}, and \texttt{blas}, then a guarded absolute path for a development cluster whose distribution omits the unversioned symbolic link; if none succeeds the build fails at configure time with an actionable message rather than at link time. \pkg{r2py\_rpart} builds a shared library loaded at import, requires no external numerical library, and is compiled as C++14 rather than C++17, because in C++17 mode the standard math header on the build system declares overloads referencing compiler built-ins that its C library does not provide; the inline-variable declarations in the reconstructed headers compile in C++14 as an accepted extension. Platform-conditional linking selects the appropriate C++ standard library on macOS.

Binary wheels are built in CI with \texttt{cibuildwheel} across five runners covering Linux \texttt{x86\_64} and \texttt{aarch64} (both \texttt{glibc} and \texttt{musl} variants), macOS \texttt{x86\_64} and \texttt{arm64}, and Windows \texttt{x86\_64}, for CPython 3.10 through 3.14, and published on release through OIDC trusted publishing. Windows builds obtain their toolchain from MSYS2, and dependent libraries are vendored into the wheel.

\subsection*{Test generation and reference comparison}

Public interfaces were identified from the package documentation; for \pkg{rpart}, cross-referencing the manual pages against the file marking internal entry points yielded 23 public functions. One sub-agent was dispatched per public function, sequentially, each reading the function's signature and the package's reference manual and emitting three categories of test. \emph{Positive} tests exercise the documented combinations of parameters and check the returned values against R. \emph{Negative} tests exercise every explicit error condition in the R source and check that the Python function refuses the same inputs. \emph{Boundary} tests exercise the edges of the input domain---empty and single-element inputs, non-finite values, degenerate ranges, and the parameter values at which the algorithm changes behavior.

Reference values were obtained from the original R packages through \pkg{rpy2}~\cite{gautier2008rpy2}. R packages are attached once at module level; inputs are passed as R vectors; results are extracted by name and converted to NumPy arrays; and agreement is asserted elementwise. Relative tolerances for assertions that compare output against R, each written at the assertion or inherited from its comparison routine, range from $10^{-10}$ to $10^{-3}$ in \pkg{r2py\_kernsmooth} and from $10^{-12}$ to $10^{-4}$ in \pkg{r2py\_rpart} (Table~\ref{tab:twopkg}); the median assertion is $10^{-6}$ in the former and $10^{-7}$ in the latter. An absolute floor is added where references can be zero. In \pkg{r2py\_rpart} the tightest bounds apply to quantities read directly from the fitted tree---every assertion at $10^{-12}$ is in the \texttt{meanvar} suite---and the loosest to cross-validated quantities: all four assertions at $10^{-4}$ are in the \texttt{xpred.rpart} suite. Fold membership is supplied identically to both implementations, so these bounds absorb error accumulated in refitting and combining across folds, not a differing partition. In \pkg{r2py\_kernsmooth} the tightest bounds apply to grid coordinates, at $10^{-10}$, and the two loosest agreement bounds, $10^{-3}$ and $10^{-4}$, are both two-dimensional density estimates at extreme scales or asymmetric bandwidths. Where exact row correspondence with an R dataset was required, the data were read from the R session rather than from a file. Where a suite is required to run without an R installation, reference values and datasets were captured once from a live R session into checked-in files; this was done for \pkg{rpart}'s translated historical suite.

Failures were resolved by an iterative loop: the suite is re-run, the first remaining failure is passed to a repair agent together with its traceback and the relevant source, the package is rebuilt if library code changed, and the cycle repeats until the suite passes.

\subsection*{Example-workflow translation}

For \pkg{rpart}, two complete example workflows---a regression tree and a classification tree, each comprising data loading, formula-based fitting, cross-validated pruning, and plotting---were translated from R to Python and executed in both languages, with console output and rendered figures captured for comparison. Visual comparison of the resulting figures exposed four rendering divergences that no numerical assertion tested: overlapping axis tick labels where R's graphics engine suppresses them, filled rather than hollow plot markers, split labels positioned beside rather than above their branch points, and a title overlapping the tree. Each was corrected in the plotting layer. This step is required for packages with a substantial presentation surface; \pkg{KernSmooth}, whose output is numeric throughout, did not require it.

\subsection*{Root-cause analysis of the cross-validation discrepancy}

Cross-validated error estimates for deeply grown \pkg{rpart} trees differed from R in trailing rows while shallower fits matched exactly. Four candidate explanations were eliminated with direct evidence before any code was changed. Loss of precision in the checked-in data fixture was excluded by comparing the fixture byte-for-byte against the live R dataset. Divergence between the two packages' C sources was excluded by direct file comparison. Contraction of multiply-add sequences under differing compiler flags was excluded by disassembling a minimal reproduction under all four flag combinations and observing that no fused instruction was emitted in any of them, the toolchain's baseline instruction set predating the relevant extension. Optimization level and target architecture more generally were excluded by rebuilding the extension with the original package's exact compiler flags and observing no change.

Identical instrumentation was then added to both implementations, first reporting per-fold prediction errors, then candidate split values in exact hexadecimal floating-point form, and finally the inputs to and outcome of the per-node stopping test. Every logged field matched bit for bit except one: the minimum-split threshold read 20 under R and 21 in the port, constant across nodes and folds. The cause was in the control-parameter defaults, where a three-branch computation reassigned one parameter before a later branch tested whether that parameter had originally been supplied; the test therefore always saw an assigned value and executed a back-derivation intended only for the unsupplied case. Capturing the original absence in a separate flag before reassignment resolved it, and was verified against R across all four combinations of supplied and omitted arguments.

The investigation also surfaced a genuine defect in \pkg{rpart}'s own C source, initially credited with the discrepancy and later ruled out as its cause. The \texttt{oops:} fallback branch of \texttt{rundown.c}, reached when a held-out observation cannot be routed past a split---its split variable and every surrogate missing---and the majority-direction fallback is not in force, back-fills one output array correctly but leaves the corresponding entries of a second holding stale values from an earlier observation, and then writes a single element one position past that second array's range. The three arrays share one allocation, in consecutive slices, so that stray write lands on the first entry of the prediction array itself. The defect is real and present identically in both packages, but once tree shapes agreed it produces identical residue in both and therefore no observable difference; the full cross-validation table matched R after the control-parameter fix alone, with no change to \texttt{rundown.c}.

\section*{Declarations}

\subsection*{Data and code availability}

The converted Python packages are distributed on the Python Package Index (PyPI) as \pkg{r2py\_kernsmooth} and \pkg{r2py\_rpart}. Their sources, together with all intermediate artifacts of the conversion---structural-analysis JSON, language-dependency catalogs and translation guides, the reconstructed C API headers, and the complete test suites---are available at \url{https://github.com/r2py-project/python-KernSmooth} and \url{https://github.com/r2py-project/python-rpart}, with the packages additionally mirrored as standalone repositories at \url{https://github.com/r2py-project/r2py_kernsmooth} and \url{https://github.com/r2py-project/r2py_rpart}. The archived snapshots corresponding to this manuscript are deposited at Zenodo under DOIs \href{https://doi.org/10.5281/zenodo.21970781}{10.5281/zenodo.21970781} (\pkg{KernSmooth}) and \href{https://doi.org/10.5281/zenodo.21970837}{10.5281/zenodo.21970837} (\pkg{rpart}). The skill and sub-agent specifications that drive every phase are included in those repositories and reproduced in the Supplementary Information. The original R packages, \pkg{KernSmooth}~2.23-26 (Unlimited license) and \pkg{rpart}~4.1.27 (GPL-2 $\vert$ GPL-3), are available from CRAN. The converted packages follow the terms of their originals: \pkg{r2py\_kernsmooth} is distributed under the Unlimited license, and \pkg{r2py\_rpart} under GPL-2, since it reuses the original GPL-licensed C source. We ask that published work using the converted packages cite the original R package as well as this one.

\subsection*{Author contributions}
J.L. conceived and supervised the study and selected the case-study packages. Y.C. and J.L. developed the methodology. Y.C. implemented the framework, developed the software, and performed the validation experiments. Y.C. and J.L. wrote and revised the manuscript. Both authors approved the final version.

\subsection*{Competing interests}

The authors declare no competing interests.

\subsection*{Acknowledgments}

This research was supported in part by the University of Notre Dame's Center for Research Computing through the use of CRC computational resources.

\bibliographystyle{unsrtnat}
\bibliography{reference}

\clearpage

\section*{Tables}
\begin{table}[H]
	\centering\small
	\begin{tabular}{lll}
		\toprule
		& \pkg{KernSmooth} $\rightarrow$ \pkg{r2py\_kernsmooth}
		& \pkg{rpart} $\rightarrow$ \pkg{r2py\_rpart} \\
		\midrule
		R source            & 16 functions, 1 file      & 47 functions, 36 files \\
		Compiled core (as shipped by R) & Fortran~77, 11 files & C, 35 files \\
		Foreign interface   & \texttt{.Fortran()}       & \texttt{.Call()} (\texttt{SEXP}) \\
		Prologue phases     & 0                         & 5 \\
		C API symbols       & 0                         & 51 \\
		Base-R constructs   & 46                        & 172 \\
		Public functions tested & 7                     & 23 \\
		Tests               & 518                       & 846 \\
		Outcome             & 518 pass                  & 844 pass, 2 \texttt{xfail} \\
		Relative tolerance vs R (median) & $10^{-10}$--$10^{-3}$ ($10^{-6}$) & $10^{-12}$--$10^{-4}$ ($10^{-7}$) \\
		\bottomrule
	\end{tabular}
	\caption{\textbf{Two conversions spanning the foreign-function-interface axis.} Both target packages are \emph{recommended} R packages with no dependencies on contributed CRAN packages. They differ in compiled language, in the interface through which R reaches that compiled code, and in scale. Numerical agreement is asserted elementwise at a relative tolerance written at each assertion or inherited from its routine; the range spans every assertion compared against R, median in parentheses. Eight further assertions in \pkg{r2py\_kernsmooth} bound properties of the plug-in bandwidth selectors---convergence in grid size, scale invariance, monotonicity in level, agreement between scale estimators---rather than agreement with R, carry the loosest bounds and are excluded. The two \texttt{xfail} results record a single structural divergence described in the text.}
	\label{tab:twopkg}
\end{table}

\begin{table}[H]
	\centering\small
	\begin{tabular}{p{0.30\textwidth}p{0.33\textwidth}p{0.28\textwidth}}
		\toprule
		\textbf{Defect} & \textbf{Observable effect} & \textbf{Detected by} \\
		\midrule
		Marshaling layer coerced Fortran-ordered matrices back to C order on the way
		into the compiled core &
		Every multi-row, multi-column input silently transposed on entry; affects all
		matrix arguments of the fit routine &
		Static audit \\
		\midrule
		\pkg{f2py} absorbs Fortran dimension arguments; wrapper still supplies them &
		Every later argument shifted by one position; one routine crashed, four
		returned wrong values silently &
		Dynamic execution \\
		\addlinespace
		Pruning mutated its argument in place, violating R's copy-on-modify semantics & Cached arrays inconsistent with the pruned tree; heap corruption on a later cross-validation call & Dynamic execution \\
		\midrule
		Response matrix flattened in row-major rather than column-major order &
		Multi-column responses silently mispaired; invisible to any single-column test &
		Package's own historical tests \\
		\addlinespace
		Non-symmetric loss matrix flattened in the wrong order &
		Loss matrix silently transposed; wrong predicted class at the root node &
		Package's own historical tests \\
		\addlinespace
		Sentinel overwritten before it was tested, changing a control-parameter
		default from 20 to 21 &
		Divergent tree shape at nodes of exactly 20 observations; downstream
		cross-validation error resembling floating-point noise & Package's own historical tests \\
		\midrule
		Fortran scalar outputs bound by value rather than by reference &
		Three estimates always returned zero, producing a division by zero and a
		\texttt{NaN} bandwidth downstream &
		Generated test suite \\
		\addlinespace
		R's lazy default argument evaluated before, rather than after, the input is
		trimmed &
		Bandwidth computed from the untrimmed range; plausible but incorrect values &
		Generated test suite \\
		\bottomrule
	\end{tabular}
	\caption{\textbf{Representative defects and the method that found them.}
		Each entry produced numerically wrong output, or memory corruption, without raising an error at the point of the fault. Rows are grouped by the verification method that first exposed the defect, in the order the four methods were applied: static audit, dynamic execution, the package's own historical tests, then the generated suite. No method found all of them, and each found defects its predecessors had missed.}
	\label{tab:defects}
\end{table}

\clearpage

\section*{Figures}

\begin{figure}[H]
\centering
\includegraphics[width=\textwidth]{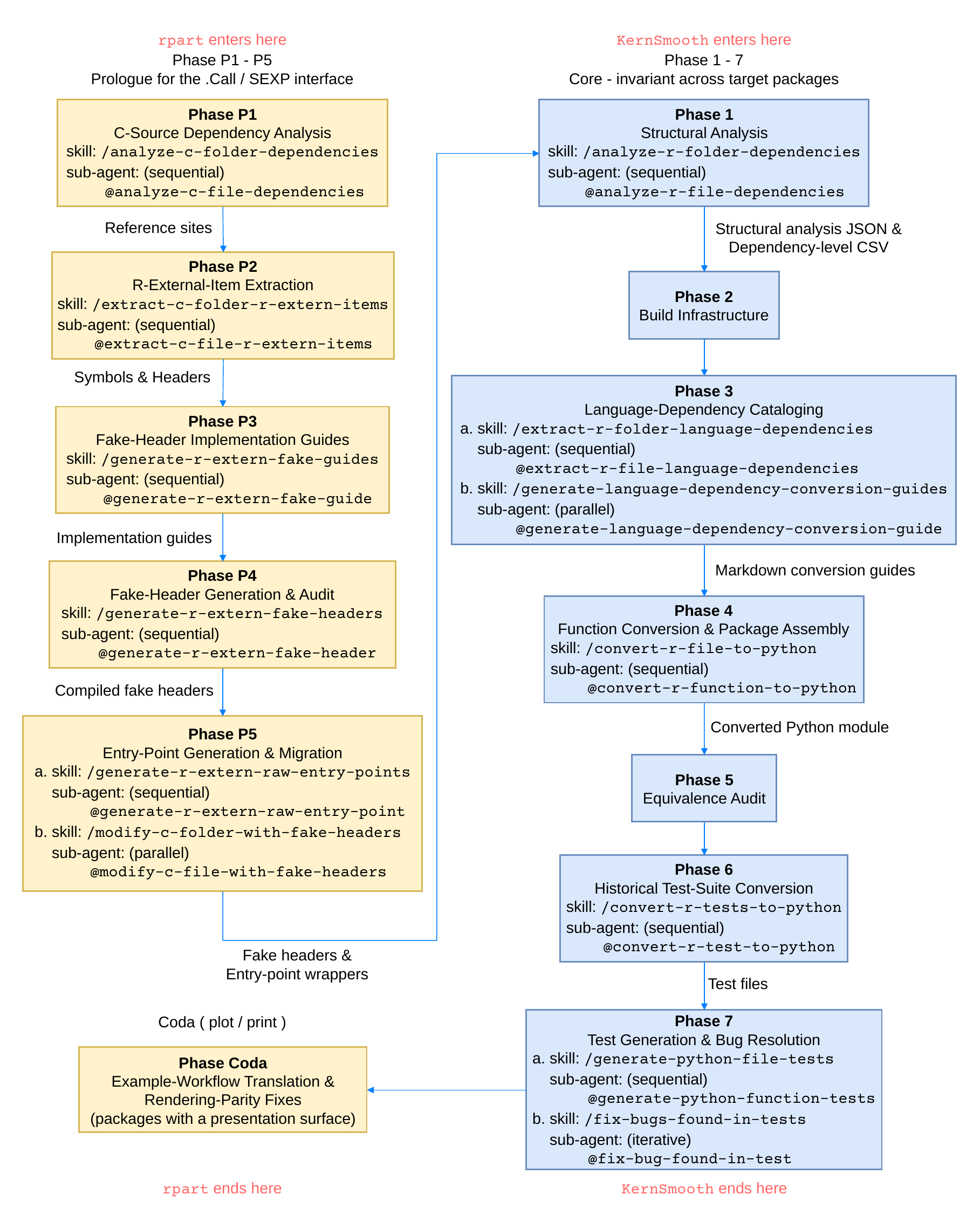}
\end{figure}

\clearpage

\captionof{figure}{\textbf{The \pkg{r2py} framework.} The core (right) carries a package from its R source to a validated Python library in seven phases: (1)~structural analysis, recovering the dependency graph over the package's own functions; (2)~build infrastructure, so that converted code is executable as soon as it is written; (3)~language-dependency cataloging, fixing the Python rendering of every base-R construct the package uses before any code is generated; (4)~function conversion and package assembly in dependency order; (5)~equivalence audit, by reading and by execution; (6)~conversion of the package's own historical test suite; and (7)~systematic test generation with iterative bug resolution. Packages whose compiled code is reached through R's array-based \texttt{.Fortran()} or \texttt{.C()} interfaces enter at phase~1, as \pkg{KernSmooth} does. Packages reached through \texttt{.Call()}, which exchanges R's own internal objects rather than plain arrays, first pass through the prologue (left), which reconstructs the portion of R's C API the package uses so that its original compiled source can be built and executed without \texttt{libR.so}; \pkg{rpart} enters there, and its fake headers and entry-point wrappers feed the core. Prologue phases are labeled P1--P5 so that the plain numerals 1--7 always denote core phases, which are invariant across packages. Whether the prologue is needed is decided by the interface; how much work it involves is set by the number of C API symbols the package references, which a static scan establishes before conversion begins. Each phase that a skill orchestrates names that skill and the sub-agent it dispatches, together with the dispatch mode: \emph{sequential} where one invocation must read an earlier one's output, \emph{parallel} where invocations are independent, and \emph{iterative} where a fix-and-recheck cycle runs to a stopping criterion. The coda is an additional step required only by packages with a substantial presentation surface; it is not part of the core, and \pkg{KernSmooth} did not require it.}
\label{fig:pipeline}

\clearpage

\clearpage

\clearpage

\end{document}